\documentclass[twocolumn]{jpsj2} 

\newcommand{\bra}[1]{\langle#1|}                  
\newcommand{\ket}[1]{|#1\rangle}                  

\newcommand{\be}{\begin{equation}}
\newcommand{\ee}{\end{equation}}
\newcommand{\bea}{\begin{eqnarray}}
\newcommand{\eea}{\end{eqnarray}}
\newcommand{\beas}{\begin{eqnarray*}}
\newcommand{\eeas}{\end{eqnarray*}}

\newcommand{\fat}[1]{\mbox{\boldmath$#1$\unboldmath}}         
\newcommand{\ham}{\hat{H}}                          
\newcommand{\dm}{\hat{\rho}}                        
\newcommand{\imag}{\text{i}}                        
\newcommand{\id}{\mathbb{I}}                        
\newcommand{\BEA}{\begin{eqnarray}}
\newcommand{\EEA}{\end{eqnarray}}
\newcommand{\rar}{\rightarrow} 
 
\title{Time-dependent Density-Matrix Renormalization-Group Methods}

\author{\textsc{Ulrich Schollw\"{o}ck}\thanks{E-mail address: scholl@physik.rwth-aachen.de}}

\inst{Institute for Theoretical Physics C, RWTH Aachen, 52056 Aachen, Germany}

\abst{In this paper recent substantial progress in applying the density-matrix 
renormalization-group (DMRG) to the simulation of the time-evolution of
strongly correlated quantum systems in one dimension is reviewed. Various
approaches to generating a time-evolution numerically are considered. The key
focus of this review is on current strategies to circumvent the 
limitations of the quite small subspace well approximated by DMRG, by either
enlarging or changing it as time evolves. All these approaches can be extended
to the simulation of mixed, i.e.\ finite-temperature states. While this 
paper is phrased almost entirely in standard DMRG language, I finish by 
considering the alternative formulation of time-evolutions in the language
of matrix product states which is less well-known but conceptually 
more powerful.}

\kword{density-matrix renormalization-group, strongly correlated systems,
quantum information theory}

\begin{document}
\maketitle

\section{Introduction}
\label{sec:intro}

More than ten years after the invention of the Density-Matrix Renormalization
Group (DMRG) by Steve White\cite{Whit92,Whit93}, 
this method has become the method
of choice for the numerical simulation of the equilibrium properties of
strongly correlated one-dimensional quantum systems\cite{Pesc99,Scho04}.
Ground-state properties of both bosonic and fermionic systems have been 
calculated often at almost machine precision and comparatively low 
computational cost. While these results have been of prime importance in
understanding the details of gapped Haldane systems or critical Luttinger 
liquids, to name but a few applications,
for a long time hardly any attention had been paid to the 
time-evolution of strongly correlated quantum systems, both due to the 
comparative lack of experimental input in the past and to the inherent
difficulties of actually calculating such time-evolutions. The last years 
have seen an increasing number of experimental results on non-trivial 
time-evolutions. Perhaps most spectacular was the recent mastery of 
storing ultracold bosonic atoms in a magnetic trap superimposed by an 
optical lattice: This has allowed to drive these atoms, at will, 
by time-dependent 
variations of the optical lattice strength,  
from the superfluid (metallic) to the Mott insulating regime. These regimes
are linked by one 
of the key phase transitions in strongly correlated systems 
\cite{Grei02}. Quite generally, progress in the fields of 
nanoelectronics and spintronics raises the question how (strongly 
correlated) quantum many-body systems react to external 
time-dependent perturbations and how transport can be calculated
quantitatively also far from the linear-response regime.
On the computational physicists' side, 
the last twelve months or so have seen an extraordinary surge of activity
in developing DMRG variants applicable to time-evolutions, such that by 
now we have various very powerful and conceptually innovative DMRG algorithms 
for pure as well as mixed quantum states, for non-dissipative as well as 
dissipative dynamics.

The relationship between these various algorithms
is often extremely close, and it is not always immediately obvious where 
the differences are of a fundamental or of a superficial nature. My goal in 
this paper is to present these algorithms {\em ahistorically} and to proceed as 
follows: after outlining the fundamental questions and difficulties, I first
show that all mixed states (i.e.\ density matrices) can be cast into the form 
of pure states such that the entire discussion of this review can focus
on the time-evolution of pure states. I move on to discussing various ways of
actually generating time-evolution generated by Hamiltonians living in
state spaces by far too large for exact diagonalization. This discussion is 
independent of DMRG specifics and serves to illustrate that in many of the
proposals there is some freedom as to the choice of the actual time-evolution
generator. DMRG takes center stage in the next section where various DMRG 
strategies of finding suitable state (sub)spaces to operate on are discussed: 
for all systems of interest here, the full state space is by far too large for
numerical treatment. Recently, it has been shown that essentially all of the 
above can be expressed in a conceptually very beautiful form in the language
of matrix product states and that computational performance and precision might
be increased that way. I will attempt to clarify the connection between 
matrix product state and DMRG time-evolution algorithms. Throughout this paper,
I assume familiarity with the basics of infinite-system and finite-system
DMRG\cite{Whit93,Scho04,Pesc99}. For notational brevity, I will consider
{\em time-independent} Hamiltonians in the formulae; 
almost all results carry over
trivially to {\em time-dependent} 
Hamiltonians except where mentioned.

Essentially all physical quantities of interest involving time can be reduced 
to the calculation of either {\em equal-time} $n$-point correlators such as the
(1-point) density
\begin{equation}
  \langle n_i(t) \rangle = \bra{\psi(t)} n_i \ket{\psi(t)} = 
  \bra{\psi} e^{\imag\ham t} n_i e^{-\imag \ham t} \ket{\psi}
\label{eq:equaltime}
\end{equation} 
or {\em unequal-time} $n$-point correlators such as the (2-point) real-time
Green's function
\begin{equation}
    G_{ij} (t) = 
\bra{\psi} c^\dagger_{i}(t) c_{j}(0) \ket{\psi} = \bra{\psi} e^{+\imag\ham 
    t} c_{i}^\dagger e^{-\imag \ham t} c_{j} \ket{\psi} .
\label{eq:unequaltime}
\end{equation}    
This expression can be cast in a form very close to Eq.\ 
(\ref{eq:equaltime}) by introducing $\ket{\phi}=c_{j}\ket{\psi}$ such that
the desired correlator is then simply given as an equal-time matrix element
between two time-evolved states,
\begin{equation}
    G_{ij}(t) = \bra{\psi(t)} c_{i}^\dagger \ket{\phi(t)} .
\label{eq:matrixelement}
\end{equation}
If both $\ket{\psi(t)}$ and $\ket{\phi(t)}$ can be calculated, a very 
appealing feature of this approach is that $G_{ij}(t)$ can be evaluated
in a single calculation for all $i$ and $t$ as time proceeds. 
Frequency-momentum space is then reached by a double Fourier 
transformation. Obviously, finite system-sizes and edge effects as well
as algorithmic constraints will impose physical 
constraints on the largest times and distances $|i-j|$ or minimal 
frequency and wave vectors resolutions accessible. 
Nevertheless, this approach might
emerge as a very attractive alternative to the current very time-consuming
calculations of $G(k,\omega)$ using the dynamical DMRG\cite{Kuhn99,Jeck02}. 

The fundamental difficulty of obtaining the above correlators become
obvious if we examine the time-evolution of the quantum state 
$\ket{\psi(t=0)}$ under the action of some 
time-independent Hamiltonian $\ham\ket{\psi_{n}}=E_{n}\ket{\psi_{n}}$.
If the eigenstates $\ket{\psi_{n}}$ are known, expanding $\ket{\psi(t=0)}=
\sum_{n}c_{n}\ket{\psi_{n}}$ leads to the well-known time evolution
\begin{equation}
    \ket{\psi(t)}=\sum_{n}c_{n}\exp (-\imag E_{n}t) \ket{\psi_{n}} , 
\label{eq:simpletime}
\end{equation}
where the modulus of the expansion coefficients of $\ket{\psi(t)}$ is 
time-independent. Now normally we are forced to restrict ourselves to some
smaller state space, as relevant Hilbert space dimensions exceed 
computational resources. A sensible Hilbert space truncation is given by a 
projection onto the large-modulus eigenstates. In strongly correlated 
systems, however, we usually have no good knowledge of the 
eigenstates. Instead, one uses some orthonormal basis with unknown 
eigenbasis expansion, $\ket{m}=\sum_{n} a_{mn} \ket{\psi_{n}}$. The 
time evolution of the state $\ket{\psi(t=0)}=
\sum_{m}d_{m}(0)\ket{m}$ then reads
\begin{equation}
    \ket{\psi(t)}=\sum_{m}\left( \sum_{n} d_{m}(0) a_{mn} e^{-\imag 
    E_{n}t} \right) \ket{m} \equiv \sum_{m} d_{m}(t) \ket{m} , 
\label{eq:complicatedtime}
\end{equation}
where the modulus of the expansion coefficients $d_m(t)$ is {\em 
time-dependent}. For a general orthonormal basis, 
Hilbert space truncation at one 
fixed time (i.e. $t=0$) will therefore not ensure a reliable 
approximation of the time evolution. Also, 
energy {\em differences} matter in time evolution due to the
phase factors $e^{-\imag (E_{n}-E_{n'})t}$ in $|d_{m}(t)|^2$. 
Thus, the sometimes justified hope that
DMRG yields a good approximation to the low-energy Hamiltonian is 
of limited use. 

\section{Time-evolution of mixed states}
\label{sec:mixed}
After the previous discussion on the difficulties of simulating the 
time-evolution of pure states in subsets of large Hilbert spaces it may seem 
that the 
time-evolution of mixed states (density matrices) is completely out of
reach. It is however easy to see that a thermal density matrix 
$\hat{\rho}_\beta \equiv \exp [-\beta \ham]$ can be constructed as a pure
state in an enlarged Hilbert space and that Hamiltonian dynamics
of the density matrix can be calculated considering just this pure state
(dissipative dynamics being more complicated).
In the DMRG context, this has first been
pointed out by Verstraete, Garcia-R\'{\i}poll and Cirac\cite{Vers04a} and
Zwolak and Vidal\cite{Zwol04}, 
using essentially 
information-theoretical language; it has also been used previously in pure
statistical physics language in e.g.\ high-temperature series 
expansions\cite{Buhl00}.

To this end, consider the completely
mixed state $\hat{\rho}_0 \equiv \id$. Let us assume that the dimension
of the local physical state space $\{\ket{\sigma_i}\}$ of a physical site 
is $N$.
Introduce now a local auxiliary state space $\{\ket{\tau_i}\}$ of the same
dimension $N$ on an auxiliary site. 
The local physical site is thus replaced by a rung of two sites, and 
a one-dimensional chain by a two-leg ladder of physical and auxiliary sites
on top and bottom rungs. Prepare now each rung $i$ in the Bell state
\begin{equation}
  \ket{\psi_0^i} = \frac{1}{\sqrt{N}} 
  \left[ \sum_{\sigma_i=\tau_i}^N |\sigma_i\tau_i\rangle \right] .
\label{eq:maxentstate}
\end{equation}
Other choices of $\ket{\psi_0^i}$ are equally feasible, as long as they
maintain in their product states maximal entanglement between
physical states $\ket{\sigma_i}$ and auxiliary states $\ket{\tau_i}$. 
Evaluating now the expectation value of some
local operator $\hat{O}_\sigma^i$ acting on the physical state space with
respect to $\ket{\psi_0^i}$, one finds
\[
\bra{\psi_0^i} \hat{O}_\sigma^i \ket{\psi_0^i} =
\sum_{\sigma_i=\tau_i} \sum_{\sigma_i'=\tau_i'} 
\frac{1}{N} \left[ \bra{\sigma_i\tau_i} \hat{O}_\sigma^i \otimes \id_\tau^i 
\ket{\sigma_i'\tau_i'} \right]. 
\]
The double sum collapses to
\[
\bra{\psi_0^i} \hat{O}_\sigma^i \ket{\psi_0^i} = 
\frac{1}{N} \sum_{\sigma_i=1}^N 
\bra{\sigma_i} \hat{O}_\sigma^i  
\ket{\sigma_i},
\]
and we see that the expectation value of $\hat{O}_\sigma^i$ with respect to
the pure state $\ket{\psi_0^i}$ living on the product of physical and 
auxiliary space is identical 
to the expectation 
value of $\hat{O}_\sigma^i$ with
respect to the completely mixed local physical state, or
\begin{equation}
\langle \hat{O}_\sigma^i \rangle = \text{Tr}_\sigma \hat{\rho}_0^i 
\hat{O}_\sigma^i
\end{equation}
where
\begin{equation}
\hat{\rho}_0^i = \text{Tr}_{\tau} \ket{\psi_0^i} \bra{\psi_0^i}.
\label{eq:rungpurificationi}
\end{equation}
This generalizes from rung to ladder using the density operator
\begin{equation}
\hat{\rho}_0 = \text{Tr}_\mathbf{\tau} \ket{\psi_0} \bra{\psi_0},
\label{eq:rungpurification}
\end{equation}
where 
\begin{equation}
\ket{\psi_0}=\prod_{i=1}^L \ket{\psi_0^i} 
\end{equation}
is the product of all local Bell states, and the conversion from ficticious
pure state to physical mixed state is achieved by tracing out all auxiliary
degrees of freedom. 

At finite temperatures $\beta>0$ one uses
\[
\hat{\rho}_\beta = e^{-\beta\ham/2} \id e^{-\beta\ham/2} = 
\text{Tr}_\mathbf{\tau} e^{-\beta\ham/2} \ket{\psi_0} \bra{\psi_0} 
e^{-\beta\ham/2},
\]
where we have used Eq.\ (\ref{eq:rungpurification}) and the observation that 
the trace can be pulled out as it acts on the auxiliary space and 
$e^{-\beta\ham/2}$ on the real space. Hence,
\begin{equation}
\hat{\rho}_\beta = \text{Tr}_\mathbf{\tau} \ket{\psi_\beta} \bra{\psi_\beta},
\label{eq:rungpurification1}
\end{equation}
where $\ket{\psi_\beta}=e^{-\beta\ham/2} \ket{\psi_0}$. Similarly, this
finite-temperature density matrix can now be evolved in time by
considering $\ket{\psi_\beta (t)}=e^{-i\ham t} \ket{\psi_\beta (0)}$ and
$\hat{\rho}_\beta(t) = \text{Tr}_\mathbf{\tau} \ket{\psi_\beta(t)} 
\bra{\psi_\beta(t)}$. The calculation of the finite-temperature time-dependent
properties of, say, a Hubbard chain, therefore corresponds to the 
imaginary-time and real-time evolution of a Hubbard ladder prepared to be
in a product of special rung states. Time evolutions generated by Hamiltonians 
act on the physical 
leg of the ladder only. As for the evaluation of expectation values
both local and auxiliary degrees of freedom are traced on the same footing, 
the distinction can be completely dropped but for the time-evolution itself.
Code-reusage is thus almost trivial. 
Note also that the initial 
infinite-temperature pure state needs only $M=1$ block states to be described
exactly in DMRG as it is a product state of single local states. Imaginary-time
evolution (lowering the temperature) will introduce entanglement such that to
maintain some desired DMRG precision $M$ will have to be increased.

\section{Generating time-evolutions in large state spaces}
\label{sec:generators}
To generate time-evolutions in large state spaces, several approaches
are feasible to integrate the Schr\"{o}dinger equation
\begin{equation}
    i\frac{\partial}{\partial t} \ket{\psi(t)} =
    [\ham - E] \ket{\psi(t)} .
    \label{eq:Schroedinger}
\end{equation}
Here, $E=\bra{\psi(0)} \ham \ket{\psi(0)}$. If $\ham$ itself is 
time-independent, energy is conserved and for reasons of numerical
efficiency a trivial phase factor of $\exp (-\imag E t)$ may be extracted
from $\ket{\psi(t)}$. In the following I assume energy shifts to set $E=0$.

On the one hand, one may now directly integrate the differential equation
forward in time, choosing some infinitesimal step size $\Delta t$. 
To lowest order in $\Delta t$, the solution is
\begin{equation}
    \ket{\psi(t+\Delta t)} = (1- \imag 
    \ham(t) \Delta t) \ket{\psi(t)}.     
    \label{eq:nonunitaryevolution}
\end{equation}
Many improvements on this simple-minded approach exist. 
In the fourth-order Runge-Kutta algorithm\cite{Caza02,Luo03,Dale04,Whit04}, 
which is perhaps the most
popular, one calculates, starting from
$\ket{\psi(t)}$, the following four {\em Runge-Kutta vectors},
\begin{eqnarray}
\ket{k_1} &=& -\imag \Delta t \ham \ket{\psi(t)}, \label{eq:rk1} \\
\ket{k_2} &=& -\imag \Delta t \ham (\ket{\psi(t)}+\frac{1}{2}\ket{k_1}), \\
\ket{k_3} &=& -\imag \Delta t \ham (\ket{\psi(t)}+\frac{1}{2}\ket{k_2}), \\
\ket{k_4} &=& -\imag \Delta t \ham (\ket{\psi(t)}+\ket{k_3}). \label{eq:rk4}
\end{eqnarray}
The wave function $\ket{\psi(t+\Delta t)}$ is then given by 
\begin{equation}
\ket{\psi(t+\Delta t)} = \ket{\psi(t)} + \frac{1}{6}\ket{k_1} +
 \frac{1}{3}\ket{k_2} + \frac{1}{3}\ket{k_3} + \frac{1}{6}\ket{k_4} .
\label{eq:rkfinal}
\end{equation} 
Note that numerical efficiency hinges centrally on the fast evaluation
of matrix-vector products $\ham \ket{\psi(t)}$.

It is important to note that Runge-Kutta does not preserve 
unitarity. This can be improved by using e.g.\ the unitary 
Crank-Nicholson time-evolution \cite{Dale04}, in its most simple version
given as
\begin{equation}
    \ket{\psi(t+\Delta t)} = \frac{1- \imag 
    \ham(t) \Delta t/2}{1+ \imag 
    \ham(t) \Delta t/2} \ket{\psi(t)}.     
    \label{eq:unitaryevolution}
\end{equation}
The drawback of this approach is that the evaluation of the non-hermitian 
denominator 
is non-trivial. This may be done by considering the large linear equation
system
\begin{equation}
(1 + \imag \ham(t) \Delta t/2) \ket{\phi} = \ket{\psi(t)},
\label{eq:sparsesystem}
\end{equation}
such that 
\begin{equation}
\ket{\psi(t+\Delta t)} = (1- \imag \ham(t) \Delta t/2) \ket{\phi}.
\end{equation}
The non-hermitian equation system (\ref{eq:sparsesystem}) may be solved
by using a biconjugate gradient approach. In fact, like all conjugate
gradient approaches, its convergence can be greatly improved by preconditioning
it, which here means the provision of the solution of some closely related,
exactly solvable equation system. Such a solution is trivially given here
by considering the diagonal ``equation system'' $\ket{\phi} = \ket{\psi(t)}$,
as $\Delta t \rightarrow 0$ ensures that Eq.\ (\ref{eq:sparsesystem}) is
diagonally dominated.

Instead of considering time-evolution as a differential equation to be
integrated numerically, one may also consider\cite{Schm04} for time-independent
Hamiltonians the formal solution of Eq.\ (\ref{eq:Schroedinger}), the
time-evolution operator $\exp (-\imag\ham t)$. This 
avoids the numerical delicacies of the integration of the Schr\"{o}dinger 
equation, but introduces new ones. It is known that the numerical evaluation of
matrix exponentials is non-trivial\cite{Mole03}; 
for example, it is {\em not} a good
strategy to consider their Taylor expansion. It seems that among the most
efficient approaches is the so-called Krylov subspace approximation.
Remembering that we are never interested in $\exp (-\imag\ham t)$, but in
$\exp (-\imag\ham t) \ket{\psi}$, one may exploit that in DMRG 
$\ham\ket{\psi}$ must be available efficiently, and form the
Krylov space by successive Schmidt-Gram orthonormalization of the
set $\{ \ket{\psi}, -\imag\ham t\ket{\psi}, (-\imag\ham t)^2\ket{\psi}, \ldots \}$ ($\ket{\psi}$
assumed normalized). As in the closely related Lanczos algorithm, the $n$ 
Krylov vectors thus obtained form the orthonormal columns of a matrix 
$V$ such that
$-\imag\ham t$ is approximated regarding its extreme eigenvalues by $VTV^T$, 
where $T$ is an $n \times n$ tridiagonal matrix, to excellent precision even for relatively
small $n$. The exponential is then given by the first column of
$V \exp T$, where the latter exponential is now much easier to calculate. 

Another way of evaluating the above matrix exponential that has its origin
in Quantum Monte Carlo methods is based on the Suzuki-Trotter 
decomposition\cite{Trot59,Suzu76}.
Its main usefulness is for Hamiltonians with nearest-neighbor interactions.
In that case we can separate interactions on odd bonds from interactions on
even bonds, such that the decomposition reads $\ham=\hat{H}_{1}+\hat{H}_{2}$;
$\hat{H}_{1}=\sum_{i=1}^{N/2} \hat{h}_{2i-1}$ and 
$\hat{H}_{2}=\sum_{i=1}^{N/2} \hat{h}_{2i}$. Here, $\hat{h}_{i}$ is the 
local Hamiltonian linking sites $i$ and $i+1$. As neighbouring local 
Hamiltonians will in general not commute, we have 
$[\hat{H}_{1},\hat{H}_{2}]\neq 0$. However, all terms 
in $\hat{H}_{1}$ or $\hat{H}_{2}$ commute. The first-order Trotter 
decomposition of the infinitesimal time-evolution operator then reads
\begin{equation}
    \exp (- \imag \ham \Delta t) = \exp (-\imag \ham_{1} \Delta t)
    \exp (-\imag \ham_{2} \Delta t) + O(\Delta t^2) .
    \label{eq:firstorderTrotter}
\end{equation}    
The second-order Trotter decomposition reads
\begin{eqnarray}
    & & \exp (- \imag \ham \Delta t) = \label{eq:secondorderTrotter} \\
    & & e^{-\imag \ham_{1} \Delta t/2}
    e^{-\imag \ham_{2} \Delta t} e^{-\imag \ham_{1} \Delta t/2} 
+ O(\Delta t^3) . \nonumber
\end{eqnarray}    
A frequently used forth-order Trotter decomposition reads
\begin{eqnarray}
    & & \exp (- \imag \ham \Delta t) = \label{eq:forthorderTrotter} \\
    & & \prod_{i=1}^5 
    e^{-\imag p_i \ham_{1} \Delta t/2}
    e^{-\imag p_i \ham_{2} \Delta t} e^{-\imag p_i \ham_{1} \Delta t/2} 
+ O(\Delta t^5) , \nonumber
\end{eqnarray}    
where all $p_i = 1/(4-4^{1/3})$, except $p_3=1-4p_1<0$, corresponding to 
evolution backward in time. 
 
For a $n$th-order Trotter decomposition, the error made in one time step 
$\Delta t$ is of order $\Delta t^{n+1}$. 
To reach a given time $t$ one has to perform 
$t/\Delta t$ time-steps, such that the error grows (at worst) {\em linearly} 
in time $t$ and
the resulting error is bounded by an expression of order $(\Delta t)^n t$. 

\section{DMRG state space selection under time-evolution}
\label{sec:selection}

All time-evolution schemes for DMRG so far can be classified into three
categories as regards their state space selection. 
{\em Static} Hilbert space DMRG methods  keep the truncated Hilbert
space found optimal for $\ket{\psi(t=0)}$. What I will call
{\em dynamic} Hilbert space DMRG methods try to enlarge 
the truncated Hilbert space starting from the one optimal to 
approximate $\ket{\psi(t=0)}$, 
such that it is big enough to 
accommodate (i.e.\ maintain large overlap with the exact result) 
$\ket{\psi(t)}$ for a sufficiently long time to a very good 
approximation. More recently, {\em adaptive} Hilbert space DMRG methods 
keep the size of the truncated Hilbert space fixed, but try to change it as 
time evolves such that it also accommodates $\ket{\psi(t)}$ to a very good 
approximation. 

\subsection{Static time-dependent DMRG}
\label{subsec:static}
 
Cazalilla and Marston\cite{Caza02} were the first to exploit DMRG to
systematically calculate time-dependent quantum many-body effects.
They studied a time-dependent Hamiltonian 
$\ham(t)\equiv \ham(0) + \hat{V}(t)$, where $\hat{V}(t)$ encodes the 
time-dependent part of the Hamiltonian.
After applying a standard DMRG calculation to the Hamiltonian 
$\ham(t=0)$, the time-dependent Schr\"{o}dinger equation was 
numerically integrated forward in time. The effective Hamiltonian
in the reduced Hilbert space was built as $\ham_{\text{eff}}(t)=
\ham_{\text{eff}}(0)+\hat{V}_{\text{eff}}(t)$, where 
$\ham_{\text{eff}}(0)$ was taken as the last superblock Hamiltonian 
approximating $\ham(0)$. $\hat{V}_{\text{eff}}(t)$ as an approximation to 
$\hat{V}$ was built using the representations of operators in the final block 
bases. The initial condition was obviously to take 
$\ket{\psi(0)}$ as the ground state obtained by the preliminary DMRG run.
This procedure amounts to fixing the reduced Hilbert space at that optimal
at $t=0$, and projecting all wave functions and operators onto it.

As an application, Cazalilla and Marston have considered a quantum dot 
weakly coupled to noninteracting leads of spinless fermions,
where time-dependency is introduced through a time-dependent chemical 
potential coupling to the number of particles left and right of the 
dot, 
\begin{equation}
    \hat{V}(t) = - \delta\mu_{R}(t) \hat{N}_{R}- \delta\mu_{L}(t) 
    \hat{N}_{L} ,
    \label{eq:bias}
\end{equation}
which is switched on smoothly at $t=0$. Setting 
$\delta\mu_{L}=-\delta\mu_{R}$, one may gauge-transform this chemical 
potential away into a time-dependent complex hopping from and to the 
dot,
\begin{equation}
    t_{q}(t) = t_{q} \exp \left[ \imag \int_{-\infty}^t dt' \delta\mu_{L}(t') 
    \right] .
    \label{eq:timehopping}
\end{equation}
The current is then given by evaluating the imaginary part of the local hopping 
expectation value. Obviously, in a finite system currents will not 
stay at some steady state value, but go to zero on a time scale of 
the inverse system size, when lead depletion has occurred. The calculation
was helped by the fact that the time-dependent coupling is in the center of 
the system which is modelled exactly by explicit sites in DMRG.

In this approach the hope is that an effective Hamiltonian 
obtained by targeting the ground state of the $t=0$ Hamiltonian is 
capable to catch the states that will be visited by the 
time-dependent Hamiltonian during time evolution. This approach
must however break down after relatively short times as the full Hilbert
space is explored.

\subsection{Dynamic time-dependent DMRG}
\label{subsec:dynamic}

Several attempts have been made to enlarge the reduced Hilbert space using
information on the time-evolution, such that the time-evolving state has
large support on that Hilbert space for longer times. Whatever procedure
for enlargement is used, the problem remains that the number of DMRG states
$M$ grows with the desired simulation time as they have to encode more and
more different physical states. As calculation time scales as $M^3$, this 
type of approach will meet its limitations somewhat later in time. 

All enlargement procedures rest on the ability of DMRG to describe -- at some
numerical expense -- small sets of states (``target states'') very well 
instead of just one. This is achieved by considering for the DMRG state
selection reduced density matrices formed as sums of some suitably chosen
pure state projectors,
\begin{equation}
\dm = \text{Tr}_E \ket{\psi}\bra{\psi} \rightarrow \dm = \text{Tr}_E \sum_i
\alpha_i \ket{\psi_i}\bra{\psi_i} , 
\end{equation}
$\sum_i \alpha_i = 1$ for normalized $\ket{\psi_i}$. $\text{Tr}_E$ is the
standard DMRG trace over the environment.

The simplest approach\cite{Dale04} 
is to target the set $\{\ket{\psi_i}\} = \{
\ket{\psi(0)}, \ham\ket{\psi(0)}, \ham^2\ket{\psi(0)}, \ldots\}$. 
Alternatively,
one might consider the Krylov vectors formed from this set. Results 
improve, but not decisively.

A much more time-consuming, but also much more performing 
approach has been demonstrated by Luo, Xiang and Wang 
\cite{Luo03}. They use 
a density matrix that is given by a superposition of states 
$\ket{\psi(t_{i})}$ at various times of the evolution,
$\dm = \sum_{i=0}^{N_{t}} \alpha_{i} 
\ket{\psi(t_{i})}\bra{\psi(t_{i})}$
with $\sum \alpha_{i}=1$
for the determination of the reduced Hilbert space.
Of course, these states    
are not known initially; it was proposed by them to
start within the framework of infinite-system DMRG from a small DMRG system and 
evolve it in time. For a very small system this procedure is exact.
For this system size, the state 
vectors $\ket{\psi(t_{i})}$ are used to form the density matrix. This
density matrix then determines the reduced 
Hilbert space for the next larger system, taking into account how 
time-evolution explores the Hilbert space for the smaller system.
One then moves on to the 
next larger DMRG system where the procedure is repeated. This is of course
very time-consuming, and it would also be of advantage to extend the
procedure to the usage of the finite-system DMRG method. 

Schmitteckert\cite{Schm04} has computed the transport through a small 
interacting nanostructure using an Hilbert space enlarging approach,
based on the time evolution operator. To this end, 
he splits the problem into two parts: By obtaining a relatively large number of 
low-lying eigenstates exactly (within time-independent DMRG 
precision), one can calculate their time evolution exactly. 
For the subspace orthogonal to these eigenstates, he implements 
the matrix exponential $\ket{\psi(t+\Delta t)}=\exp (-\imag\ham \Delta 
t)\ket{\psi(t)}$ using the Krylov subspace approximation. For any block-site
configuration during sweeping, he evolves the state in time, obtaining
$\ket{\psi(t_{i})}$ at fixed times $t_{i}$. These are targeted in the
density matrix, such that upon sweeping forth and back a Hilbert space
suitable to describe all of them at good precision should be obtained.
For numerical efficiency, he carries out this procedure to convergence
for some small time, which is then increased upon sweeping, bringing
more and more states $\ket{\psi(t_{i})}$ into the density matrix. 

\subsection{Adaptive time-dependent DMRG}
\label{subsec:adaptive} 

All approaches mentioned so far imply that to maintain good precision over
time the size of the reduced Hilbert space has to grow with desired simulation
time, imposing huge algorithmic cost. 
Time-dependent DMRG using {\em adaptive} Hilbert spaces, that ``follow'' the
state to be approximated optimally, has first been 
proposed in essentially identical form by Daley, Kollath, Schollw\"{o}ck and
Vidal\cite{Dale04} and White and Feiguin \cite{Whit04}. 
Both approaches are efficient implementations of an 
algorithm for the classical simulation of the time evolution of 
weakly entangled quantum states invented by Vidal\cite{Vida03a,Vida04} 
[time-evolving block-decimation (TEBD) algorithm]. The TEBD algorithm 
was originally formulated in the matrix product state language. For 
simplicity, I will explain the algorithm in its DMRG context; a 
detailed discussion of the very strong connection between adaptive 
time-dependent 
DMRG and the original simulation algorithm is given by Daley {\em et al.}
\cite{Dale04}: it 
turns out that DMRG naturally attaches good quantum numbers to state 
spaces used by the TEBD algorithm, allowing for the usual drastic 
increases in performance due to the use of good quantum numbers.   

For each block-site configuration 
(system-sites-environment: S$\bullet\bullet$E) 
during DMRG calculations, the truncation
is always carried out at the position of the sites; hence it can react to
changes of the physical state there. On the other hand, as these sites
are explicit, we can carry out time-evolution on this bond even exactly.
Time evolution in the adaptive time-dependent DMRG is therefore generated using 
a Suzuki-Trotter decomposition\cite{Suzu76}: 
Expanding $\ham_{1}$ and $\ham_{2}$ into the local Hamiltonians, one 
infinitesimal time step $t\rightarrow t +\Delta t$ may be carried out (taking
the most simple case of a first-order Suzuki-Trotter decomposition) 
by performing the local time-evolution on (say) all even bonds first 
and then all odd bonds. We may therefore carry out an exact time-evolution by 
performing one finite-system sweep forward and backward 
through an entire one-dimensional chain, with time-evolutions on all 
even bonds on the forward sweep and all odd bonds on the backward 
sweep, at the price of the Trotter error. This 
procedure necessitates that $\ket{\psi(t)}$ is available in the right 
block bases for the current block-site configuration, 
which is ensured by carrying out the reduced basis 
transformations on $\ket{\psi(t)}$ that in standard DMRG form the 
basis of White's prediction method \cite{Whit96b}. The decisive idea of 
Vidal \cite{Vida03a,Vida04} in the TEBD algorithm was now to carry out a new 
Schmidt decomposition (in DMRG language: reduced density matrix formation) 
and make a new choice 
of the most relevant block basis states for $\ket{\psi(t)}$ 
after each local bond update. Therefore, as the quantum state changes 
in time, so do the block bases such that an optimal representation of 
the time-evolving state is ensured. Choosing new block bases changes 
the effective Hilbert space, hence the name {\em adaptive} 
time-dependent DMRG. 

An appealing feature of this algorithm is that it can be very easily 
implemented in existing finite-system DMRG. One uses standard 
finite-system DMRG to generate a high-precision initial state 
$\ket{\psi(0)}$ and continues to run finite-system sweeps, one for 
each infinitesimal time step, merely replacing the large 
sparse-matrix diagonalization at each step of the sweep by local bond 
updates for the odd and even bonds, respectively. 

An extensive error discussion has been carried out by Gobert {\em et al.}
\cite{Gobe04}. Two main sources of errors occur, the Trotter error due
to the Suzuki-Trotter decomposition, and the truncation error due to
the loss of information about the quantum state at each DMRG projection
onto a reduced basis. As discussed above, the Trotter error scales
with time as $O(\Delta t^n t)$ for a $n$th order Suzuki-Trotter 
decomposition. Of course, there are prefactors associated depending on
the details of the decomposition. In the setup of Gobert {\em et al.}
the error scales (maximally) 
linearly with system size $L$, and overall it is thus of 
order $(\Delta t)^n L t$. 
Further errors are produced by the truncations (reduced basis 
transformations) after each bond update during an infinitesimal 
time step. While the truncation error $\epsilon$ that sets the scale of
the error of the wave function and operators is typically very small in
DMRG, here it will strongly accumulate over time
as $O(Lt/\Delta t)$ truncations are carried out up to time $t$. This is because
the truncated DMRG wave function has norm less than one and is renormalized
at each truncation by a factor of $(1-\epsilon)^{-1}>1$. Truncation errors
should therefore accumulate roughly exponentially with an exponent
of $\epsilon Lt/\Delta t$, such that 
eventually the adaptive t-DMRG will break down at too long times.
In practice, saturation effects in errors and partial compensations of errors
in observables will lead to somewhat slower error growth. The very different 
growth behaviour of these errors leads to a well-defined ``runaway time'',
at which the Trotter error is replaced by the accumulated truncation error
as dominant error. It can be seen in a sharp onset of error growth for
physical observables, either be comparison to higher-precision (larger M)
calculations or to exact solutions, when available.
The errors mentioned can be well 
controlled by increasing $M$, the runaway time, which signals that
adaptive t-DMRG will break down in the imminent future, growing
linearly\cite{Gobe04} with $M$.  

Applications of adaptive time-dependent DMRG so far for the 
(time-dependent) Bose-Hubbard model \cite{Dale04,Koll04}, spin-1 Heisenberg 
chains \cite{Whit04}, and far-from-equilibrium states in
spin-1/2 Heisenberg chains\cite{Gobe04} 
have demonstrated that it allows to access 
large time scales (several hundred in units of inverse interaction
energy or bandwidth) very reliably. 

Adaptive t-DMRG based on Suzuki-Trotter decompositions is for all practical 
purposes restricted to nearest-neighbour interactions.
In a second adaptive approach without that limitation, 
White and Feiguin\cite{Whit04b,Whit04c} exploit that there is no 
conceptual need in DMRG to combine the actual time-evolution and the adaptation
of the state spaces into one step. In fact, even the time-scales $\delta t$ 
on which time-evolution is discretized and $\Delta t$ 
on which state spaces are adapted need not be identical: the empirical 
observation that even a static state space remains a good choice for some 
finite time, allows to choose $\Delta t \gg \delta t$. The actual 
time-evolution may be carried out using any of the non-Trotter methods
described above; White and Feiguin have chosen fourth-order Runge-Kutta
integration, which is a valid and convenient, but perhaps not the ultimate 
choice. One may alternatively use, for example, Crank-Nicholson integration or
Krylov-vector based exponentials\cite{Manm2004,Frie2004}.

To adapt the state space, at intervals $\Delta t$, an analysis is carried
out on the states that will appear in the near future. To that purpose,
several DMRG sweeps are carried out at a fixed time $t$. For each block-site
configuration during these sweeps, a Runge-Kutta integration up to
$t+\Delta t$ is carried out. The reduced density matrix that during sweeping
controls the choice of the reduced block bases is formed from the present 
state and the states representing the near future. Here resides an ambiguity.
While the Runge-Kutta vectors $\ket{k_i}$ encode time-evolution up to
$t+\Delta t$, the density matrix choice $\dm_S = \text{Tr}_E 
(\ket{\psi(t)}\bra{\psi(t)}+ \sum_i \ket{k_i}\bra{k_i})$ is not found to be 
optimal. The Runge-Kutta vectors are essentially encoding derivatives
of state vectors, while what one wants to approximate are state vectors at 
times up to $t+\Delta t$. The density matrix during sweeping is therefore 
constructed as
\begin{equation}
  \dm_S = \text{Tr}_E \sum_{i=1}^4 \ket{\psi(t_i)}\bra{\psi(t_i)}
\label{eq:rkdensity}
\end{equation}
where $t_1=t$, $t_2=t+\Delta t/3$, $t_3=t+2\Delta t/3$,
$t_4=t+\Delta t$. Here, of course modifications are conceivable. 
The state vectors at the intermediate times $t+\Delta t/3$ and $t+\Delta 2t/3$ 
are given by
\begin{eqnarray}
& & \ket{\psi(t+\Delta t/3)} = \ket{\psi(t)}+  \\
& & \frac{1}{162}\left[ 31\ket{k_1}+14(\ket{k_2}+\ket{k_3})-5\ket{k_4} \right]
\nonumber \\
& & \ket{\psi(t+\Delta 2t/3)}= \ket{\psi(t)}+ \\
& & \frac{1}{81} \left[ 16\ket{k_1}+20(\ket{k_2}+\ket{k_3})-2\ket{k_4} \right]
\nonumber
\end{eqnarray}
They can be calculated from exploiting that Runge-Kutta methods are exact for 
polynomials up to some degree, which establishes a connection between
the Runge-Kutta vectors and the state at all intermediate times. 

This approach is free of the Trotter error, but the error analysis of the
truncation error should remain valid. In fact, there might be for long-time
simulations additional errors from finding the approximate Hamiltonian
(i.e.\ the reduced state space on which the full Hamiltonian is projected)
by carrying out a time-evolution using just that approximate Hamiltonian. That 
there is some such error can be seen from the observation that for $\Delta t
\rightarrow \infty$ one essentially recovers static time-dependent DMRG with 
its limitations. Of course, for any {\em finite} $\Delta t$, 
the adaptive DMRG will be (much) better than static DMRG. $\Delta t$ 
therefore has to be chosen optimally: small enough to update the Hilbert space 
often enough, large enough to avoid too much numerical overhead.
At the time of writing, a complete error analysis still remains to be done. 
The price to be paid for the elimination of the Trotter error and the
generalization to longer-ranged interactions 
is that during the finite-system sweeps at intervals of
$\Delta t$, a substantial number of time-consuming $\ham\ket{\psi}$ 
multiplications have to be
carried out, which can be completely avoided in the Trotter-based approach.

As the time-evolution of density matrices can be simulated by using pure
states in enlarged state spaces, all the above carries over to the 
simulation of imaginary time-evolutions. Two comments are in order.
(i) The completely mixed state can be described in DMRG using $M=1$. After 
a very long imaginary time-evolution, it will reach the ground state,
for which -- given years of ground state DMRG experience -- substantial
values of $M$ (typically hundreds) have to be chosen for high precision.
It is therefore a sensible approach to carry out imaginary time evolutions
fixing some truncation error and to choose $M$ at each time step such that
it is not exceeded. This will amount to keeping more and more states as
temperature is being lowered. (ii) Imaginary-time evolution is not unitary,
but reduces norms. This leads to compensating effects for the accumulation
of errors and imaginary-time evolution is numerically easier to handle than 
real-time evolution.

\section{An alternative formulation: matrix-product states}
\label{sec:mpa} 
As has been pointed out by various authors 
\cite{Ostl95,Mart96b,Romm97,Duke98,Taka99},  
the ansatz states that DMRG generates are very closely related 
to {\em position-dependent 
matrix-product states} (also known as finitely correlated states) 
\cite{Fann89,Klum93}. However, there are subtle, 
but crucial differences between DMRG states and matrix-product states 
\cite{Taka99,Vers04}. Recently, consistent use of matrix-product states
has given rise to new algorithms closely related to 
DMRG\cite{Vers04,Vers04a} that are of high relevance for 
time-evolution. Let me first state some properties of matrix-product
states. 
Matrix-product states are simple generalizations of 
product states of local states, which we take to be on a chain,
\begin{equation}
    \ket{\fat{\sigma}} = \ket{\sigma_{1}} \otimes \ket{\sigma_{2}} \otimes \ldots 
    \otimes \ket{\sigma_{L}} ,
    \label{eq:productstate}
\end{equation}
obtained by introducing linear operators $\hat{A}_{i}[\sigma_{i}]$ 
depending on the local state. These operators map from some 
$M$-dimensional auxiliary state space spanned by an orthonormal basis 
$\{\ket{\beta}\}$ to another $M$-dimensional auxiliary state space 
spanned by $\{\ket{\alpha}\}$:
\begin{equation}
\hat{A}_{i}[\sigma_{i}]=\sum_{\alpha\beta} 
(A_{i}[\sigma_{i}])_{\alpha\beta} \ket{\alpha}\bra{\beta} .
\label{eq:ansatzmatrix}
\end{equation}
One may visualize the auxiliary state spaces to be located on the 
bonds $(i,i+1)$ and $(i-1,i)$.
The operators are thus represented by
$M\times M$ matrices $(A_{i}[\sigma_{i}])_{\alpha\beta}$; $M$ will be seen 
later to be the number of block states in DMRG.
We further demand for reasons explained below that
\begin{equation}
\sum_{\sigma_{i}} \hat{A}_{i}[\sigma_{i}]\hat{A}_{i}^\dagger[\sigma_{i}] = \id .
\label{eq:mpanorm}
\end{equation}    
A position-dependent unnormalized matrix-product state for a 
one-dimensional system of size $L$ is then given by
\begin{equation}
    \ket{\psi} = \sum_{\{\fat{\sigma}\}} \left( \bra{\phi_{L}} 
    \prod_{i=1}^L 
    \hat{A}_{i} [\sigma_{i}]
    \ket{\phi_{R}} \right) \ket{\fat{\sigma}},
    \label{eq:mpsopen}
\end{equation}
where $\bra{\phi_{L}}$ and $\ket{\phi_{R}}$
are left and right boundary states in the auxiliary state spaces 
located in the above visualization to the left of the first and to the right 
of the last site. They are used to 
obtain scalar coefficients. Position-independent matrix-product states are 
obtained by making Eq.\ (\ref{eq:ansatzmatrix}) 
position-independent, $\hat{A}_{i}[\sigma_{i}]\rar \hat{A} 
[\sigma_{i}]$.
For simplicity, we shall consider only those in the following.

For periodic boundary conditions, boundary states are replaced  
by tracing the matrix-product:
\begin{equation}
    \ket{\psi} = \sum_{\{\fat{\sigma}\}} \text{Tr} \left[\prod_{i=1}^L 
    \hat{A} 
    [\sigma_{i}]
    \right] \ket{\fat{\sigma}} .
    \label{eq:mpsperiodic}
\end{equation}
The best-known matrix-product state is the valence-bond-solid ground 
state of the bilinear-biquadratic $S=1$ Affleck-Kennedy-Lieb-Tasaki 
Hamiltonian 
\cite{Affl87,Affl88}, where $M=2$.

If we consider operators $\hat{O}_{j}$ and $\hat{O}_{j+l}$ acting on sites 
$j$ and $j+l$, for a periodic boundary condition matrix-product state 
the correlator $C(l)=\bra{\psi} \hat{O}_{j}\hat{O}_{j+l} \ket{\psi} / 
\langle \psi | \psi \rangle$
is given by\cite{Ande99}
\begin{equation}
    C(l)=
    \frac{\text{Tr} \overline{O}_{j} \overline{\id}^{l-1} \overline{O}_{j+l} 
    \overline{\id}^{L-l-1}}{\text{Tr} \overline{\id}^{L}}, 
    \label{eq:mpscorrelator}
\end{equation}    
where we have used the following mapping \cite{Romm97,Ande99} from an 
operator $\hat{O}$ acting on the local state space to a
$M^2$-dimensional operator $\overline{O}$ acting on products of
auxiliary states $\ket{\beta\beta'}=\ket{\beta}\otimes \ket{\beta'}$:
\begin{equation}
    \overline{O} = \sum_{\sigma\sigma'} \bra{\sigma'} \hat{O}
   \ket{\sigma} \hat{A}^* 
    [\sigma'] \otimes \hat{A} [\sigma] .
    \label{eq:mpsmapping}
\end{equation}
Note that $\hat{A}^*$ stands for $\hat{A}$ 
complex-conjugated only as opposed to 
$\hat{A}^\dagger$. Evaluating Eq.\ (\ref{eq:mpscorrelator}) in the 
eigenbasis of the mapped identity, $\overline{\id}$, we find that in 
the thermodynamic limit $L\rightarrow\infty$
\begin{equation}
    C(l)= \sum_{i=1}^{M^2} c_{i} \left( \frac{\lambda_{i}}{|\lambda_{i}|} \right)^l
    \exp (-l/\xi_{i})
    \label{eq:decaympa}
\end{equation}    
with $\xi_{i}=-1/\ln|\lambda_{i}|$. The $\lambda_{i}$ are the 
eigenvalues of $\overline{\id}$, and the $c_{i}$ depend on $\hat{O}$.
This expression holds because due to Eq.\ 
(\ref{eq:mpanorm}) $|\lambda_{i}|\leq 1$ and 
$\lambda_{1}=1$ for the eigenstate $\langle 
\beta\beta'|\lambda_{1}\rangle = \delta_{\beta\beta'}$. Equation
(\ref{eq:mpanorm}) is thus seen to ensure normalizability of matrix 
product states in the thermodynamic limit. 
Generally, all correlations in matrix-product states are either
long-ranged or purely exponentially decaying. 

That a DMRG calculation
retaining $M$ block states produces states closely related to 
$M\times M$ matrix-product states as introduced above was first 
realized by Ostlund and Rommer\cite{Ostl95}. Let us consider 
(in the framework of finite-system DMRG) the reduced basis transformation 
to obtain the block states $\ket{m_i}$ of a 
{\em right} block including sites $i,\ldots,L$ from 
the right block states $\ket{m_{i+1}}$ of sites $i+1,\ldots,L$ and the
states $\ket{\sigma_i}$ of site $i$:
\begin{equation}
    \langle m_{i+1}\sigma_{i} | m_{i} \rangle \equiv 
    (A_{i})_{m_{i};m_{i+1}\sigma_{i}} \equiv 
     (A_{i}[\sigma_{i}])_{m_{i};m_{i+1}},
\label{eq:indexedtrafomatrix}
 \end{equation}    
such that 
\begin{equation}
    \ket{m_{i}} = \sum_{m_{i+1}\sigma_{i}} 
    (A_{i}[\sigma_{i}])_{m_{i};m_{i+1}}\ket{\sigma_{i}}\otimes\ket{m_{i+1}}.
    \label{eq:indexedtrafo}
\end{equation}    
The reduced basis transformation matrices $A_{i}[\sigma_{i}]$ 
automatically obey Eq.\ 
(\ref{eq:mpanorm}), which here ensures that $\{ \ket{m_{i}}\}$ is 
an orthonormal basis provided $\{ \ket{m_{i+1}}\}$ is one, too.

For the {\em left} block of sites $1,\ldots,i$, similarly define
\begin{equation}
    \langle m_{i-1}\sigma_{i} | m_{i} \rangle \equiv 
    (A_{i})_{m_{i-1}\sigma_{i};m_{i}} \equiv 
     (A_{i}[\sigma_{i}])_{m_{i-1};m_{i}},
\label{eq:indexedtrafomatrix1}
 \end{equation}    
which implies
\begin{equation}
\sum_{\sigma_{i}} \hat{A}^\dagger_{i}[\sigma_{i}]\hat{A}_{i}[\sigma_{i}] = \id ,
\label{eq:mpanorm1}
\end{equation}    
a variation of Eq.\ (\ref{eq:mpanorm}) that also ensures normalization. 

We may now use Eq.\ (\ref{eq:indexedtrafo}) for a backward recursion 
to express $\ket{m_{i+1}}$ via $\ket{m_{i+2}}$ and so forth. I will
show results for the right blocks, the left blocks can be treated 
analogously. 
There is a (conceptually irrelevant) complication as the number of 
block states for very short blocks is less than $M$. For simplicity, 
I assume that $N^{\tilde{N}}=M$, and stop the recursion 
at the shortest right block of size $\tilde{N}$ that has $M$ states, 
such that
\begin{eqnarray}
    \ket{m_{i}} &=& \sum_{\fat{\sigma}_{RB}}
    \sum_{\sigma_i,\ldots,\sigma_{L-\tilde{N}}} 
    (A_{i} [\sigma_{i}] \ldots A_{L-\tilde{N}} 
    [\sigma_{L-\tilde{N}}])_{m_{i}\fat{\sigma}_{RB}}\times \nonumber 
    \\
    & &   \ket{\sigma_{i}\ldots\sigma_{L-\tilde{N}}}\otimes \ket{\fat{\sigma}_{RB}},
\end{eqnarray}    
where we have boundary-site states 
$\ket{\fat{\sigma}_{RB}}\equiv\ket{\sigma_{L-\tilde{N}+1}\ldots\sigma_L}$;
hence
\begin{equation}
    \ket{m_{i}} = \sum_{\sigma_{i},\ldots,\sigma_{L}} 
    (A_{i} [\sigma_{i}] \ldots A_{L-\tilde{N}} 
    [\sigma_{L-\tilde{N}}])_{m_{i},\fat{\sigma}_{RB}} 
    \ket{\sigma_{i}\ldots\sigma_{L}} . 
    \label{eq:DMRGproductstate}
\end{equation}    
DMRG thus generates position-dependent 
$M\times M$ matrix-product states as block states for a reduced 
Hilbert space of $M$ states; the auxiliary state space to a local state 
space is given by the Hilbert space of the block to which the local site 
is the latest attachment. 
DMRG looks for the ground state of the full 
chain as the variational optimum in energy 
in a space spanned by products of two local states and two 
matrix-product states, 
\begin{equation}
\ket{\psi} = \sum_{m_{i-1} m_{i+2}} \sum_{\sigma_i \sigma_{i+1}}
\psi_{m_{i-1}\sigma_i \sigma_{i+1} m_{i+2}} 
\ket{m_{i-1}\sigma_i \sigma_{i+1} m_{i+2}}.
\label{eq:DMRGstate}
\end{equation}
Encoding the $\psi$-coefficients as a set of $M\times M$ matrices
parametrized by $\sigma_i \sigma_{i+1}$, $\Psi_{m_{i-1} m_{i+2}}
[\sigma_i \sigma_{i+1}]=
\psi_{m_{i-1}\sigma_i \sigma_{i+1} m_{i+2}}$, we have
\begin{eqnarray}
    \ket{\psi} &=& \sum_{\{\fat{\sigma}\}} 
(A_{\tilde{N}+1}[\sigma_{\tilde{N}+1}]\ldots A_{i-1} [\sigma_{i-1}] 
    \Psi[\sigma_i \sigma_{i+1}] \times \nonumber \\ & &
 A_{i+2} [\sigma_{i+2}] \ldots A_{L-\tilde{N}} 
     [\sigma_{L-\tilde{N}}])_{\fat{\sigma}_{LB},\fat{\sigma}_{RB}} 
\nonumber \\
     & & \ket{\sigma_{1}\ldots\sigma_L} . \label{eq:DMRGmeetsmpa}
\end{eqnarray}    

The effect of the finite-system DMRG algorithm can be seen from Eq.\ 
(\ref{eq:DMRGmeetsmpa}) to be a sequence of {\em local} optimization 
steps of the wave function that have two effects: on the one hand, 
the variational 
coefficients $\Psi[\sigma_i\sigma_{i+1}]$ are optimized, 
on the other hand, a new improved 
ansatz matrix $A$ is obtained for the site added into the growing block, 
using the improved variational coefficients from $\Psi$ 
for the reduced density matrix to obtain $A$.

On closer inspection, the DMRG state  to describe $\ket{\psi}$
is different from a true matrix-product 
state: There is an anomaly in  
that formal ``translational invariance'' of this state is broken by the 
indexing of $\Psi$ by {\em two} sites. This suggests to consider instead
a setup where $\Psi$ is indexed by {\em one} site:
\begin{eqnarray}
    \ket{\psi} &=& \sum_{\{\fat{\sigma}\}} 
(A_{\tilde{N}+1}[\sigma_{\tilde{N}+1}]\ldots A_{i-1} [\sigma_{i-1}] 
    \Psi[\sigma_i] \times \nonumber \\ & &
 A_{i+1} [\sigma_{i+1}] \ldots A_{L-\tilde{N}} 
     [\sigma_{L-\tilde{N}}])_{\fat{\sigma}_{LB},\fat{\sigma}_{RB}} 
\nonumber \\
     & & \ket{\sigma_{1}\ldots\sigma_L} . \label{eq:DMRGmeetsmpa1}
\end{eqnarray}    

In the block-site language of DMRG this would correspond to a
S$\bullet$E instead of an S$\bullet\bullet$E setup where bullets stand 
for sites. The (finite-system) 
DMRG algorithm for this S$\bullet$E setup can now be reformulated 
in this picture as follows: 
sweeping forward and backward through the chain, one keeps for site 
$i$ all $A$ at other sites
fixed and seeks the $\Psi_{i}$ that minimizes the total energy. 
From this, one determines $A_{i}$ and 
moves to the next site, seeking $\psi_{i+1}$ (or $\psi_{i-1}$, depending
on the direction of the sweep), and so on, until all 
matrices have converged.

The key point is that in the ansatz of Eq.\ (\ref{eq:DMRGmeetsmpa1}) the 
variational optimum possible for a matrix product state of dimension $M$,
whereas this is not the case for the ansatz of Eq.\ (\ref{eq:DMRGmeetsmpa}),
albeit it may come very close. This is the major difference between
DMRG ansatz and matrix product states. To see this consider how $A[\sigma_i]$
and the variationally optimal $\Psi[\sigma_i]$ are related in the new setup: 
$A$ is obtained from a reduced density matrix which is formed from tracing out 
the environment in the pure state projector $\ket{\psi}\bra{\psi}$.
In the new setup the environment is a block of dimension $M$. 
Linear algebra tells us 
that the maximum number of non-zero eigenvalues of the $MN$-dimensional density
matrix thus obtained is then bounded by $M$. ``Truncation'' down to the
$M$ highest-weight eigenstates of the reduced density matrix is hence an 
{\em exact} basis transformation of the variationally obtained state of minimal 
energy. There is no loss of information. Shifting the active site $i$ through 
the system means taking out some old $A$, look for the $\Psi$ with minimal 
energy at its place, and find a new $A$ from it. 
As there is a $\Psi$ corresponding to that old $A$ (i.e.\ leading to the same
energy of the state; it can be constructed explicitly), 
the minimal energy can only stay constant or go down,
such that this sequence of improvements of the ansatz matrices 
is truly variational in the 
space of the states generated by the maps $A$ and reaches a minimum 
of energy within that space (there is of course no guarantee to reach 
the global minimum). By comparison, the setup S$\bullet\bullet$E 
leads to a reduced basis transformation with generic loss of information 
as the dimension of the environment is $MN$. It is hence not strictly 
variational.

Indeed, in practical applications it has been observed that for 
translationally invariant systems with periodic boundary conditions the 
position dependency of observables is not eliminated entirely
during-finite system sweeps in standard DMRG. 
Dukelsky {\em et al.}\cite{Duke98} and Takasaki, Hikihara and 
Nishino\cite{Taka99} were the first to observe that switching, 
after convergence is reached, from the 
S$\bullet\bullet$E scheme for the finite-system DMRG algorithm to a 
S$\bullet$E scheme for further sweeps eliminates this dependency. 

Let us now turn back to the original topic of this paper.
To calculate time-evolutions in matrix-product states, 
Verstraete, Garcia-R\'{\i}poll and Cirac\cite{Vers04a} consider the 
$T=0$ matrix product state of Eq.\ (\ref{eq:mpsperiodic}). The 
$A_{i}[\sigma_{i}]$ are now interpreted as maps from 
the tensor product of the two auxiliary states (dimension $M^2$) next
to a site to the physical state space of the site of dimension $N$. 
This can be written as:
$(A_{i}[\sigma])_{\alpha\beta}$: $A_{i} = 
\sum_{\sigma_{i}}\sum_{\alpha_{i}\beta_{i}}  
(A_{i}[\sigma])_{\alpha\beta}
\ket{\sigma_{i}}\bra{\alpha_{i}\beta_{i}}$. The $\ket{\alpha}$ 
and $\ket{\beta}$ are states of the auxiliary state spaces
$a_i$ and $b_i$ on the left and right bonds emerging from site $i$.

The state of Eq.\ (\ref{eq:mpsperiodic}) and its interpretation can be 
generalized to finite-temperature matrix product {\em density 
operators}
\begin{equation}
    \hat{\rho} = \sum_{\{\fat{\sigma}\}\{\fat{\sigma}'\}} \text{Tr}
    \left[\prod_{i=1}^L \hat{M}_{i} 
	[\sigma_{i}\sigma_{i}']
	\right] \ket{\fat{\sigma}}\bra{\fat{\sigma}'} .
	\label{eq:mpdoperiodic}
\end{equation}    
Here, the $ \hat{M} [\sigma_{i}\sigma_{i}']$ now are maps from the 
tensor product of four auxiliary state spaces (two left, two right of 
the physical site) of dimension $M^4$ to the 
$N^2$-dimensional local density-operator state space. 
The most general case allowed by quantum mechanics for such mappings 
is for $\hat{M}$ to be a 
completely positive map. The dimension of $\hat{M}$ seems to be 
prohibitive for making practical use of this variational ansatz 
(\ref{eq:mpdoperiodic}), but it is known from quantum information theory that
$\hat{M}$ can be decomposed into a sum of $d$ tensor 
products of $A$-maps as
\begin{equation}
    \hat{M}_{i} [\sigma_{i}\sigma_{i}'] = \sum_{a_{i}=1}^{d}
    A_{i} [\sigma_{i}a_{i}] \otimes A_{i}^*  [\sigma_{i}'a_{i}].
    \label{eq:completemapdecomp}
\end{equation}
In general, $d\leq N M^2$, but it will be important to 
realise that for thermal states $d= N$ only. At $T=0$, 
one recovers the standard matrix product state, i.e.\ $d=1$. In order 
to actually simulate $\hat{\rho}$, Verstraete {\em et al.}\cite{Vers04a} 
consider the purification
\begin{equation}
    \ket{\psi_{MPDO}} = \sum_{\{\fat{\sigma}\}\{\fat{\tau}\}}
    \text{Tr}
	\left[\prod_{i=1}^L \hat{A}_{i} [\sigma_{i}\tau_{i}]
	    \right] \ket{\fat{\sigma}\fat{\tau}} ,
    \label{eq:purification}
\end{equation}
such that $\hat{\rho}=\text{Tr}_{\fat{\tau}} 
\ket{\psi_{MPDO}}\bra{\psi_{MPDO}}$.
Here, ancilla state spaces $\{ \ket{\tau_i} \}$ of dimension 
$d$ have been introduced. 

In this form, a completely mixed state is 
obtained from matrices $A_{i}[\sigma_{i}\tau_{i}] \propto \id \cdot 
\delta_{\sigma_{i},\tau_{i}}$, where $M$ may be 1 and normalization has 
been ignored. This shows $d= N$. This state is now 
subjected to infinitesimal 
evolutions in imaginary time, $e^{-\ham\Delta t}$, up
to the imaginary time $\beta/2$. As they act on $\sigma$ only, 
the dimension of the ancilla state 
spaces need not be increased. Of course, for $T=0$ the state may also be  
efficiently prepared using standard methods. The introduction of 
an ancilla state space
of dimension $N$ is the complete equivalent of the motivation of
the auxiliary state space introduced in Sec.\ \ref{sec:mixed}.

The imaginary-time evolution is carried out after a Trotter 
decomposition into infinitesimal time steps on bonds. The local bond 
evolution operator $\hat{U}_{i,i+1}$ is conveniently decomposed into a sum of 
$d_{U}$ tensor products of on-site evolution operators,
\begin{equation}
    \hat{U}_{i,i+1} = \sum_{k=1}^{d_{U}} \hat{u}_{i}^k \otimes \hat{u}_{i+1}^k .
    \label{eq:onsitedecomp}
\end{equation}
$d_{U}$ is typically small, say 2 to 4. Applying now 
the time evolution at, say, all odd bonds exactly, the auxiliary 
state spaces are enlarged from dimension $M$ to $Md_{U}$. One now 
has to find the optimal approximation $\ket{\tilde{\psi}(t+\Delta t)}$ 
to $\ket{\psi(t+\Delta t)}$ using auxiliary state spaces of dimension 
$M$ only. Hence, the state spaces of dimension $Md_{U}$ 
must be truncated optimally to minimize $\| \ket{\tilde{\psi}(t+\Delta 
t)} - \ket{\psi(t+\Delta t)}\|$. If one uses the matrices composing 
the state at $t$ as 
initial guess and keeps all $A$-matrices but one fixed, one obtains a 
linear equation system for this $A$; sweeping through all $A$-matrices 
several times is sufficient to reach the fixed 
point which is the variational optimum. As temperature is lowered, 
$M$ will be increased to maintain a desired precision.
Once the thermal state is constructed, real-time evolutions governed 
by Hamiltonians can be 
calculated similarly. It therefore turns out that this method is essentially
identical to the Trotter-based adaptive time-dependent DMRG, with the
advantages of finding the true variational optimum and of global truncation
after one entire infinitesimal time step as opposed to local truncations
after each bond time step. Presently, it seems that the gain in precision
is not extremely significant, and studies of relative computational costs
for a desired precision are still lacking.

\section{Conclusion and Outlook}

I hope to have shown that there are now multiple efficient ways of 
simulating Hamiltonian dynamics of pure and mixed quantum states for
strongly correlated one-dimensional quantum systems in the DMRG framework.
This should open the way to a huge number of interesting new DMRG applications
and be of interest for years to come. Very little is still known about the
detailed performance of the new algorithms at the moment, but what is
known is indeed very promising. In fact, I have passed over the fact that
all the above can also extended to master equation dynamics of
density matrices, i.e.\ dissipative dynamics\cite{Vers04a,Zwol04} -- even 
less is known here at the moment. But algorithmically, everything is
very close to the non-dissipative case:
The approach of Zwolak and Vidal\cite{Zwol04} is based on the 
observation that local 
(density) operators $\sum \rho_{\sigma\sigma'} \ket{\sigma}\bra{\sigma'}$ 
can be represented as $N\times N$
hermitian matrices. They now reinterpret the matrix coefficients as 
the coefficients of a local ``superket'' defined on a 
$N^2$-dimensional local state space, and TEBD/DMRG-style dynamics 
can be carried out on that superket. Similarly, 
Verstraete {\em et al.}\cite{Vers04a} consider dissipative dynamics for
states of the form of Eq.\ (\ref{eq:mpdoperiodic}),
which is formally equivalent to a matrix product state on a local 
$N^2$ dimensional state space. As should be clear 
from the discussion of matrix product states, both approaches are 
effectively identical but for the optimal approximation of 
Verstraete {\em et al.} 
\cite{Vers04a} replacing the somewhat less precise truncation of 
the approach of Zwolak and Vidal\cite{Zwol04}. Needless to say, even more
interesting applications are waiting here! 

\section*{Acknowledgment}
I would like to thank the organizers of the STATPHYS satellite meeting 
SPQS 2004 in Sendai, Japan, foremost Prof.\ Seiji Miyashita and Dr. Synge Todo, 
for all their efforts and hospitality.

\end{document}